\documentclass[12pt]{article} 
\usepackage{amssymb}

\newtheorem{thm}{Theorem}
\newtheorem{cor}{Corollary}

\newcommand{\proof}{\noindent {\em Proof:\/}\ }  

\newcommand{\qed} {\hfill \rule{0.1in}{0.1in}} 

\newcommand\reals{\mathbb R}

\def\comp{\mathop{\rm comp}\nolimits}

\def\il{\left\langle}
\def\ir{\right\rangle}
\def\eps{\varepsilon}
\def\e{\varepsilon}

\def\c{{\bf c}}
\def\l{\lambda}

\def\a{\alpha}

\def\qq{\mbox{qq}(\e,F_r)}

\title{Path Integration on a Quantum Computer\thanks{This research was
supported in part by the National Science Foundation. Effort 
sponsored by the Defense Advanced Research Projects Agency
(DARPA) and Air Force Research Laboratory, Air Force Materiel Command,
USAF, under agreement number F30602-01-2-0523. The U.S, Government is
authorized to reproduce and distribute reprints for Governmental 
purposes notwithstanding any copyright annotation thereon.
The views and conclusions contained herein are those of the authors and
should not be interpreted as necessarily representing the official policies or
endorsements, either expressed or implied, of the Defense Advanced
Research Projects Agency (DARPA), the Air Force Research Laboratory,
or the U.S. Government.}}
\author{
J. F. Traub\\ { Computer Science, Columbia University}\\
H. Wo\'zniakowski\\
 { Computer Science, Columbia University and}\\
{ Institute of Applied Mathematics, University of Warsaw}}

\begin{document}

\maketitle

\begin{abstract}
We study path integration on a quantum computer that performs
quantum summation. We assume that the measure of path integration 
is Gaussian, with the eigenvalues of its covariance operator 
of order $j^{-k}$ with $k>1$.
For the Wiener measure occurring in many  applications
we have $k=2$. We want to compute an $\e$-approximation
to path integrals whose integrands are at least Lipschitz. We prove: 
\begin{itemize}
\item Path integration on a quantum computer is tractable.

\item Path integration on a quantum computer can be solved roughly
      $\e^{-1}$ times faster than on a classical computer using 
      randomization, and exponentially faster than on a classical computer
      with a worst case assurance. 
\item The number of quantum queries needed to solve path integration
      is roughly the square root of the number of function values 
      needed on a classical computer using randomization. More
      precisely, the number of quantum queries is at most $4.22\,
      \e^{-1}$. Furthermore, a lower bound is
      obtained for the minimal number of quantum queries which shows
      that this bound cannot be significantly improved.

\item The number of qubits is polynomial in $\e^{-1}$.
      Furthermore, for the Wiener measure the degree is $2$
      for Lipschitz functions, and the degree is $1$ for 
      smoother integrands. 
\end{itemize}
\end{abstract}

\section{Introduction}

Although quantum computers currently exist only as prototypes 
in the laboratory, we believe it is important
to study theoretical aspects of quantum computation and to investigate 
its potential power. There will be additional incentives to try to
build quantum computers if it can be shown that 
there are substantial speed-ups for a variety of problems. 

To date there have been two major algorithms for discrete
problems on quantum computers that are significantly better than on 
classical computers: Shor's factorization and Grover's data search 
algorithms, see \cite{G96,G98,S94,S98}. 
But numerous problems in science and engineering 
have  {\it continuous} mathematical models. 
Examples include high dimensional integrals, path integrals, 
partial differential and integral
equations, and continuous optimization.

Continuous problems are usually solved numerically; they can
only be solved to within uncertainty $\e$. The computational
complexity of these problems on classical computers is often known;
for a recent survey see \cite{TW98}. Complexity is defined to be
the minimal number of function values and arithmetic operations
needed to solve the problem to within~$\e$.

For many continuous problems defined on functions of $d$ variables,
the complexity in the worst case deterministic setting is 
exponential in $\e^{-1}$ or in $d$. In the latter case,
the problem is said to suffer from  the ``curse of dimensionality'' and
is computationally intractable.
For some continuous problems the curse of dimensionality can be vanquished
by weakening the worst case deterministic assurance to a
stochastic assurance, such as in the randomized setting.
Monte Carlo is a prime example of an algorithm in the randomized setting.

A start has been made toward solving continuous problems on 
quantum computers in recent papers \cite{AW99, H01a,H01b,HNa,HNb,N01,NSW}.
They study multivariate integration and approximation.
The major technical tool in these papers is the quantum summation
algorithm of Brassard, H\o yer, Mosca and Tapp that is based on 
Grover's iterate, see \cite{BHMT00,G96}. 
The essence of the results of Heinrich and Novak, \cite{H01b,HNa,N01},
is that intractability in the worst case setting of
multivariate integration in a Sobolev space is {\em broken} by the use
of the quantum summation algorithm. That is, we have an {\em exponential}
speed-up of quantum algorithms over deterministic algorithms 
with a worst case assurance. Furthermore, there is roughly a {\em quadratic}
speed-up of quantum algorithms over randomized algorithms run 
on a classical computer. 

Our paper is a continuation of the idea of using quantum summation 
for continuous problems. Summation is often required for
continuous problems. Algorithms such as Monte Carlo and Quasi-Monte
Carlo are used for a variety of continuous problems and they require
the summation of many terms. In the worst case  setting, 
the number of terms $n$ is often an exponential function of $\e^{-1}$.
However, if we perform summation on a quantum computer, 
this is {\em not} a show stopper, since the cost of the quantum 
summation algorithm depends 
only logarithmically on $n$. Hence, as long as $n$ is a single exponential
function of $\e^{-1}$ the quantum cost is {\em polynomial},
and the problem becomes {\em tractable} on a quantum computer. 
In this paper we show that quantum summation is a powerful tool for
computing also path integrals. 

Path integrals may be viewed as integration of functions 
of {\em infinitely} many variables. They occur in many 
fields, including quantum
physics and chemistry, differential equations, and financial mathematics.
Efficient algorithms for approximating path integrals are therefore
of great interest. However, and perhaps not surprisingly, 
path integration is {\it intractable} on a classical computer
in the worst case setting for 
integrands with finite smoothness as shown in \cite{WW96}. 
Fortunately, the worst case complexity of path integration is only a single
exponential function in $\e^{-1}$ if the measure of path integration
is Gaussian and the eigenvalues of the covariance operator are of order
$j^{-k}$ for $k>1$. For the Wiener measure, which appears in many applications,
we have $k=2$. That is why when we use the quantum summation algorithm, 
path integration becomes {\it tractable} on a quantum computer. 
More precisely, for functions having smoothness $r$,
see the precise definition of the class $F_r$ in Section 3, 
path integrals can be computed using of order 
\begin{itemize}
\item $\e^{-1}$ quantum queries,
\item $\e^{-(k+\gamma(r))/(k-1)}\,\log \e^{-1}$ quantum operations, and 
\item $\e^{-(1+\gamma(r))/(k-1)}\,\log\,\e^{-1}$ qubits.
\end{itemize}
Here $\gamma(1)=1$ and $\gamma(r)=0$ for $r\ge2$. 
For the Wiener measure, we have more specific bounds, which we 
present in Theorem 2. We stress that the cost of a quantum query
depends on a particular applications and may be very large. 

We know more precise bounds on the number of quantum queries.
To explain them we comment on two types of errors for
path integration. The first error occurs when we replace the original 
problem by finite dimensional Gaussian integration, and then the
second one when we approximate the finite dimensional problem
by a finite sum. For simplicity, we assume that the both errors are
bounded by $\e/2$ so that the total error is at most $\e$. 
Hence, we need to apply the quantum summation algorithm with error $\e/2$.    
To get a $\delta$-error, the quantum summation algorithms requires 
at most $2.11\delta^{-1}$ quantum queries and this bound is 
in general sharp, see \cite{KW}. Hence, for $\delta=\e/2$ we need
at most $4.22\e^{-1}$ quantum queries. 

Obviously, we may reduce the number of quantum queries by choosing 
a different splitting of the two errors for path integration. So, if the
first error is, say, $a\e$ and the second is $(1-a)\e$ for some $a\in
(0,1)$, the number of required quantum queries is at most
$2.11\e^{-1}/(1-a)$, and for small $a$ it is roughly at most
$2.11\e^{-1}$. This can be achieved at the expense of increasing
quantum operations and qubits.

We also study the question what is the minimal number of quantum
queries for solving path integration by an arbitrary quantum
algorithm. Similarly as in \cite{N01}, we show that path integration
is no easier than a specific summation problem. Then using the lower
bound of \cite{NW98}, we conclude that the minimal number of quantum
queries is at least of order $\e^{-1+\a}$ for any $\a\in(0,1)$, 
see Theorem 3. This means that the number of quantum
queries used by the algorithm presented in this paper 
cannot be significantly improved.

We stress that the number of qubits is polynomial in $\e^{-1}$.
Furthermore, for the Wiener measure the degree is $2$ for $r=1$,
and $1$ for $r\ge2$. Hence, if $\e$ is relatively large we do not
need too many qubits to solve path integration on a quantum computer.
This is important since the number of qubits will be a limiting resource
for the foreseeable future. 

{}From these bounds and from the known complexity bounds in the worst
case and randomized settings, we conclude that
\begin{itemize}
\item Path integration on a quantum computer can be solved roughly
      $\e^{-1}$ times faster than on a classical computer using 
      randomization, and exponentially faster than on a classical computer
      with a worst case assurance. 
\item The number of quantum queries is the square root of 
      the number of function values needed on a classical
      computer using randomization. 
\end{itemize}

We outline the remainder of this paper. In Section 2  we briefly discuss
the complexity of summation in the worst case and randomized settings,
and the quantum summation algorithm for computing the arithmetic mean
of $n$ numbers, each from the interval $[-1,1]$. In Section 3 we
define path integration precisely, 
while in Section 4 we explain a computational 
approach to path integration. In Section 5 we summarize what is known 
about the complexity of path integration on a classical computer
in the worst case and randomized settings. We also outline an 
algorithm of Curbera, \cite{C00}, which requires 
exponentially many function values in the worst case setting,  
and which is the basis for the quantum
path integration algorithm. In Section 6 we discuss 
path integration on a quantum computer
and summarize the advantages of the quantum algorithm.
In Section 7 we prove that the upper bound on the number of
quantum queries presented in Section 6 is essentially minimal.
In the Appendix we present the proof of how many variables
must be used to approximate path integrals to within $\e$. 

\section{Quantum Summation Algorithm}

Sums occur frequently in scientific computation. For example, when
Monte Carlo or Quasi-Monte Carlo are used to approximate a 
$d$-dimensional integral, we compute $n^{-1}\sum_{i=1}^nf(x_i)$,
where the $x_i$ are $d$-dimensional vectors that are chosen randomly
(for Monte Carlo) or deterministically (for Quasi-Monte Carlo),
see e.g., \cite{N92}. As we shall see in Section 5, such algorithms
can be also used for approximating path integrals. In fact, for many linear
problems it is known that linear algorithms enjoy many optimality properties,
see e.g., \cite{N88,TWW88,TW98}.
Linear algorithms have 
the form $\sum_{i=1}^na_if(x_i)$ for coefficients $a_i$ that 
are sometimes, but not always, equal to $n^{-1}$. Let $y_i=a_if(x_i)n$.
Then for all these applications we wish to compute
\begin{equation}\label{sum}
S_n(y)\,=\,n^{-1}\,\sum_{i=1}^ny_i.
\end{equation}

In this paper we restrict ourselves to the case when $|y_i|\le 1$ for 
$i=1,2,\dots,n$. 
More general conditions on the $y_i$ of the form
$\left(n^{-1}\sum_{i=1}^n|y_i|^p\right)^{1/p}\le 1$ with $p\in[1,\infty]$
are considered in \cite{H01a,H01b,HNa,HNb}.

We are interested in applications where $n$ is huge. We wish to approximate 
$S_n$ to within $\e$ for $\e\in (0,{1\over 2})$. 
The terms $y_i$ are not stored or computed in advance. We assume
that for a given index $i$ we have a subroutine that computes $y_i$. 
This assumption is typical for scientific problems where, as explained above,
$y_i$ depends on the function value $f(x_i)$. 

Before we discuss quantum
computation of $S_n$, we briefly mention summation complexity results
in the worst case and randomized settings
on a classical computer, see \cite{N88,N01}.
The worst case complexity, $\comp^{{\rm wor}}(n,\e)$, is defined as the 
minimal number of operations needed to compute an $\e$-approximation to
$S_n$ for all $|y_i|\le 1$ using deterministic algorithms. 
The randomized complexity, $\comp^{{\rm ran}}(n,\e)$, is defined 
analogously when we permit randomized algorithms. It is known that
$$
\comp^{{\rm wor}}(n,\e)\,\approx\,n\,(1-\e),
$$
and if $n\gg \e^{-2}$,
$$
\comp^{{\rm ran}}(n,\e)\,\approx\,\e^{-2}.
$$
Hence, in the worst case setting we must add essentially all $n$ numbers,
whereas in the randomized setting it is enough to add only $\e^{-2}$ terms
and this, of course,  can be achieved by the Monte Carlo algorithm
that chooses $\e^{-2}$ samples from the set $\{y_1,y_2,\dots,y_n\}$, 
each with probability $n^{-1}$, and computes  their arithmetic mean.
This speed-up is significant.

We now turn to what is known about summation on a quantum computer.
We wish to compute $QS_n(y,\e)$ which approximates $S_n(y)$ to within
$\e$ with probability at least ${3\over 4}$. That is, $QS_n(y,\e)$ is a random
variable for which the inequality $|S_n(y)-QS_n(y,\e)|\le \e$ holds
with probability at least ${3\over 4}$. The performance of a quantum algorithm
can be summarized by the number of 
quantum queries, quantum operations and qubits which are used,
see \cite{BHMT00,G98,H01a,N01} for precise definitions of  
quantum computation and quantum algorithms. 
Here, we only mention that the quantum algorithm obtains 
information on the terms $y_i$ by using only quantum queries.
The number of quantum operations is defined as the total number 
of bit operations performed by the quantum algorithm, and the 
number of qubits is defined as $k$ if all quantum computations
are performed in the Hilbert space of dimension~$2^k$. 

Since the number of qubits will be a limiting resource
for the foreseeable future, it is important to seek
algorithms which require as few qubits as possible. 

Brassard, H\o yer, Mosca and Tapp, see 
\cite{BHMT00}, presented  a quantum algorithm $QS_n$ that 
solves the summation problem. Their algorithm 
is based on Grover's iterate, see \cite{BHMT00,G98}, and uses quantum Fourier 
and Walsh-Hadamard transforms that can be implemented by well known
quantum gates. 
Assume that $n\gg \e^{-1}$.  Then the algorithm $QS_n$ uses of order 
\begin{eqnarray*}
\e^{-1}&\qquad& \mbox{quantum queries},\\
\e^{-1}\,\log\,n &\qquad& \mbox{quantum operations, }\\
\log\,n\, 
&\qquad& \mbox{qubits}. 
\end{eqnarray*}
More precise bounds are known about the number of quantum queries.
In \cite{KW}, it is shown that the quantum algorithm $QS_n$ uses
at most $\e^{-1}\,2.10\dots$ quantum queries and this bound is sharp for
small $\e$ and large $\e n$.    
Due to the lower bound of Nayak and Wu, see \cite{NW98}, the number of quantum
queries of {\em any} quantum algorithm that solves the summation problem must
be at least of order $\e^{-1}$. Hence, the algorithm $QS_n$
uses almost the minimal number of quantum queries. 
(In this paper $\log$ denotes $\log_2$.)

We can run the quantum algorithm $QS_n$ several times
to increase the probability of success. If we want to solve the
problem with probability $1-\delta$, then we should run $QS_n$ 
roughly $\log\,\delta^{-1}$ times and take the median as our 
final result. Then the number of queries and quantum operations
is multiplied by $\log\,\delta^{-1}$, but the number of qubits stays
the same. 

Of course, these quantum results are of interest only 
if $\e^{-1}$ is significantly less than $n$.
Fortunately, this is the case for a number of important problems. 
Indeed, this paper will supply
one more such problem, namely, path integration. 

So far we assumed that we summed numbers from the interval $[-1,1]$. 
The interval $[-1,1]$ is taken only for simplicity.
If we have the interval $[-M,M]$ then we can rescale the summands
to $y_i/M$, and multiply the computed result by $M$. 
This corresponds to the previous problem over the
interval $[-1,1]$ with $\e/M$. Note, however, 
that for large $M$, and $n>M/\e$,
the quantum cost is of order $M/\e$, which is significantly larger
than $1/\e$. 

\vskip 1pc
\section{Definition of Path Integration}

We now define path integrals studied in this paper, see also \cite{WW96}. 
Let $X$ be an infinite dimensional separable Banach space
equipped with a probability measure $\mu$. 
We assume that $\mu$ is a zero mean Gaussian measure, see e.g., \cite{V87}.
The space $X$ can be embedded in the Hilbert space $H=L_2([0,1])$  
for which the embedding $\mbox{Im}:X\to H$ is a continuous linear 
operator. The inner product of $H$ 
is denoted by $\il\cdot,\cdot\ir_H$. Then 
the measure $\nu\,=\,\mu\,\mbox{Im}^{-1}$ is also a zero mean Gaussian 
measure on the Hilbert space $H$. Let $C_\nu$ be the covariance
operator of $\nu$, i.e., $C_\nu:H\to H$ and
$$
\il C_\nu h_1,h_2\ir_H\,=\,\int_H\il h,h_1 \ir_H\il h,h_2 \ir_H\,\nu(dh)
\qquad \forall\,h_1,h_2\in H.
$$
The operator $C_\nu$ is self
adjoint, nonnegative definite and has a finite trace. 
We can assume that 
there exists an orthonormal system $\{\eta_i\}$ from $\mbox{Im}(X)$,
$\il\eta_i,\eta_j\ir_H\,=\,\delta_{i,j}$, for which
\[
C_\nu\,\eta_i\,=\,\l_i\,\eta_i,
\]
\begin{equation}\label{eig}
\l_1\,\ge\,\l_2\,\ge\,\cdots\,\ge\,0\quad\mbox{and}\quad
\sum_{i=1}^\infty\l_i\,<\,+\infty.
\end{equation}

We illustrate this definition by the important example of 
the space $X=C([0,1])$ of continuous functions defined on $[0,1]$
with the sup norm, $\|x\|=\max_{t\in[0,1]}|x(t)|$.
The space $C([0,1])$ is equipped with the classical Wiener
measure $\mu=w$. The measure $w$ is a zero mean Gaussian measure with
covariance function $\min(t,u)$. That is,
\begin{eqnarray*}
\int_{C([0,1])}x(t)\,w(dx)\,&=&\,0\qquad \forall\,t\in [0,1],\\
\int_{C([0,1])}x(t)\,x(u)\,w(dx)\,&=&\,\min(t,u)
\qquad \forall\,t,u\in [0,1].
\end{eqnarray*}
For the Wiener measure $w$, we have $\mbox{Im}(x)=x$ and 
\[
\eta_i\,=\,\sqrt{2}\,\sin\left(\frac{2i-1}{2}\pi x\right),\qquad 
\l_i\,\,=\,\frac{4}{\pi^2(2i-1)^2}.
\]
\vskip 1pc
We return to the case of general $X$ and $\mu$. 
Let $F$ be a class of real-valued $\mu$-integrable functions defined on $X$.
The {\em path integration} problem is defined  
as approximating integrals of $f$ from $F$,
\begin{equation}\label{int}
  I(f)\,:=\,\int_Xf(x)\,\mu(dx)\,=\,\int_Hf(\mbox{Im}^{-1}x)\,\nu(dx)
,\quad \forall\,f\in F. 
\end{equation}
If only finitely many eigenvalues $\l_i$ of $C_\nu$ are positive, then
the measure $\nu$ is concentrated on a finite dimensional subspace
of $H$ and path integration reduces to a finite dimensional 
Gaussian integration. To preserve the main feature of the path integration
problem, which is integration over an infinite dimensional space, 
we assume that all eigenvalues $\l_i$ are positive. 
The element $x$ from $H$ can be written as
$x=\sum_{i=1}^{\infty}t_i\eta_i$, with $t_i=
\il x,\eta_i\ir_H$. Therefore the integrand $f$ in (\ref{int}) 
depends on infinitely many variables $t_i$.
That is why the path integration problem can be viewed as 
integration of functions having infinitely many variables. 

In this paper we will consider the classes  
$F_r$ of functions whose $r-1$ times Frechet derivatives exist and
are bounded, and whose $r$th Frechet derivatives satisfy 
the Lipschitz condition. More precisely, for a non-negative integer
$i$, let $\|f^{(i)}\|\,=\,\sup_{x\in X}\|f^{(i)}(x)\|$.
Here, $f^{(i)}(x)$ is an $i$-linear form from $X^i$ 
to $\reals$, and its norm is defined as $\|f^{(i)}(x)\|\,=\,
\sup_{\|x_j\|_X\le1}|f^{(i)}(x)x_1x_2\,\cdots\,x_i|$.
Obviously, $\|f^{(0)}\|=\|f\|=\sup_{x\in X}|f(x)|$. 

Let $r$ be a positive integer. 
For positive numbers $K_0,K_1\dots,K_r$, define
$\|f\|_{r-1}=\max_{0\le i\le r-1}\|f^{(i)}\|/K_i$. The class 
$F_r$ is defined as 
$$
F_r\,=\,\{\,f\,:\  \|f\|_{r-1}\le 1, \
\|f^{(r-1)}(x)-f^{(r-1)}(y)\|\,\le\,K_{r}\|{\rm Im}(x-y)\|_H,\ \forall\,
x,y\in X\}.
\]

For $r=1$, the class $F_1$ consists of bounded  Lipschitz functions.
The values of $f$ are bounded by $K_0$, and the Lipschitz constant by $K_1$.
For $r\ge 2$, the class $F_r$ consists of bounded smooth functions.
All functions from $F_r$ are $r-1$ times Frechet differentiable, their
$i$th derivatives are bounded by $K_i$ for $i=0,1,\dots,r-1$,
and the $(r-1)$st derivatives satisfy the Lipschitz condition with 
the constant $K_r$. 

Note that for any 
$f\in F_r$, the path integral $I(f)$ is well defined since $f$ is 
continuous and bounded.  {}From  $|f(x)|\le K_0$ we have $|I(f)|\le K_0$. 
If $K_0$ is large we can use a different estimate on $I(f)$.
We have $I(f)=f(0)+ I(f-f(0))$ and $|I(f-f(0))|\le K_1\int_H\|x\|\nu(dx)
\le K_1\left(\sum_{j=1}^\infty\l_j\right)^{1/2}$. Hence,
$|I(f)|\le |f(0)| + K_1\left(\sum_{j=1}^\infty\l_j\right)^{1/2}$.
This estimate can be better than the previous one for large $K_0$.

As we shall see in the next sections, path integration for the class
$F_r$ is intractable in the worst case setting. We stress that for
other classes of functions, path integration can be tractable even in
the worst case setting. An example is provided for the class 
of smooth integrands occurring in the Feynman-Kac formula, see
\cite{KL,PWW}.

\section{Computational Approach to Path Integration}

We want to approximate $I(f)$ to within $\eps$ for all $f \in F$.
The approximate computation of $I(f)$ consists
of two steps, see \cite{WW96}. The first is to approximate the infinite 
dimensional integration $I$ by a $d$-dimensional 
integration $I_d$, where $d=d(\e,F)$ is chosen as the minimal
integer for which the error
of this approximation is at most, say, ${\e\over 2}$. 
The second step is to compute
an approximation to $I_d$ with error at most ${\e\over 2}$.
Clearly, we should expect that $d(\e,F)$ would go to infinity as $\e$
goes to zero.

More precisely we proceed as follows. 
Let $f_d:\reals^d\to \reals$ be defined for  
$t\,=\,[t_1,t_2,\dots,t_d]\in \reals^d$ as
\begin{equation}\label{1580}
f_d(t\,)\,=\,
f\left(\mbox{Im}^{-1}(t_1\eta_1\,+\,t_2\eta_2\,+\,\cdots\,+\,t_d\eta_d)\right).
\end{equation}
Define
\begin{equation}\label{intd}
I_d(f)\,=\,
\frac1{(2\pi)^{d/2}}\,\frac1{\sqrt{\l_1\l_2\cdots\l_d}}
\int_{\reals^d}f_d(t\,)\,\exp\left(-t_1^2/(2\l_1)-\cdots
-t_d^2/(2\l_d)\right)\ d\, t.
\end{equation}
Observe that $I_d$ is a finite dimensional Gaussian integral
with the eigenvalues $\l_i$ as variances. Note that 
the eigenvalues $\l_i$ tend to zero. Indeed, since 
$a=\sum_{i=1}^\infty\l_i\,<\,+\infty$ and $\l_i$ are non-increasing
then $\l_i\,\le\,a/i$ for all $i$. Hence, we have  
decreasing dependence on the successive variables $t_i$ in (\ref{intd}).
For continuous $f$, we have
\[
I(f)\,=\,\lim_dI_d(f_d). 
\]
As outlined above, we want to choose the minimal $d=d(\e,F)$ such that
$|I(f)-I_d(f)|\le {\e\over 2}$ $\forall f\in F$,  and then to 
compute an ${\e\over 2}$-approximation to a finite-dimensional 
integral $I_d(f_d)$. 
We now find $d(\e,F_r)$ for a family of eigenvalues $\l_j$
of the covariance operator $C_\nu$. The family includes the eigenvalues 
of $C_\nu$ for the Wiener measure. 
\vskip 1pc
\begin{thm}\label{thm-1}
Suppose  $\l_j$  is of order $j^{-k}$ with $k>1$. Then 
\begin{eqnarray*}
\underline{c}_1\,\e^{-2/(k-1)}\,\le\,d(\e,F_{1})\,&\le&\,
\overline{c}_1\,\e^{-2/(k-1)},\\
\underline{c}_r\,\e^{-1/(k-1)}\,\le\,d(\e,F_{r})\,&\le&\,
\overline{c}_r\,\e^{-1/(k-1)}
 \ \  \mbox{for}\ r\ge 2,
\end{eqnarray*}
where $\underline{c}_r$ and $\overline{c}_r$, for $r=1,2,\dots$, are
positive numbers independent of $\e$ and depending only on 
the global parameters
$K_i,r,k$ and the trace $\sum_{i=1}^\infty\l_i$. 
In particular, if $\l_j=aj^{-k}$ with $a>0$ then 
\begin{eqnarray*}
d(\e,F_1)\,&\le&\,1\,+\,
\left(\frac{4aK_1^2}{k-1}\right)^{1/(k-1)}\,\left(\frac1{\e}
\right)^{2/(k-1)},\\
d(\e,F_r)\,&\le&\,1\,+\,
\left(\frac{aK_2}{k-1}\right)^{1/(k-1)}\,\left(\frac1{\e}\right)^{1/(k-1)}
\qquad \mbox{for}\ \ r\ge2. 
\end{eqnarray*}
For the Wiener measure, $\l_j=4/(\pi^2(2i-1)^2)$, we have 
\begin{eqnarray*}
d(\e,F_1)\,&\le&\, \bigg\lceil\left(\frac1{\pi^2}\, 
\frac{2K_1}{\e}\right)^2\,+\frac12\bigg\rceil,\\
d(\e,F_r)\,&\le&\,\bigg \lceil \frac{K_2}{\pi^2\e}\,+\,\frac12\bigg \rceil
\qquad \mbox{for}\ \ r\ge2. 
\end{eqnarray*}
\end{thm}
\vskip 1pc
The proof of this theorem is given in the Appendix.
We stress that the upper bounds on $d(\e,F_r)$ in Theorem 1 
depend only on $K_1$ for $r=1$, and on $K_2$ for $r\ge2$,  
i.e., on the Lipschitz constants for $f$ or $f^{\prime}$, respectively. 
This means that we can even take all the remaining $K_i=\infty$ 
and the upper bounds on $d(\e,F_r)$ still hold. 
On the other hand, the lower bounds 
depend on all of the $K_i$. The dependence is weak since they only
affect the multiplicative factors of the power of $\e^{-1}$, and the power
of $\e^{-1}$ does not depend on $K_i$.

\section{Path Integration on a Classical Computer}

In this section we discuss approximation of path integrals on a 
classical computer. We assume the {\it real number} 
model of computation, which is usually used for 
the analysis of scientific computing problems,
see \cite{TPT} for the rationale. 
We assume, in particular, that  
we can perform arithmetic operations (addition, subtraction,
multiplication, division), and comparisons of real
numbers. We assume that these operations are performed exactly and 
each costs unity. To approximate path integrals
we must have information concerning the integrands $f\in F$. 
This information may be supplied by function values $f(x_i)$ 
for some $x_1,x_2,\dots,x_n$,
where $n=n(\e,F)$ will be chosen depending on the error
demand $\e$ and the class $F$.
As outlined in the previous section,
we will need to know $f(x)$ for $x$ belonging to a finite dimensional 
subspace $X_d=\mbox{span}(\mbox{Im}^{-1}\eta_1,\mbox{Im}^{-1}\eta_2,
\dots,\mbox{Im}^{-1}\eta_d)$ with $d=d(\e,F)$. 
We therefore assume that we can compute values of $f(x)$ for
$x\in X_d$ and the cost of one such evaluation is $\c_d$.
Usually $\c_d\gg 1$. Furthermore, we will sometimes assume that the cost
$\c_d$ depends linearly on $d$, i.e., $\c_d\,=\,\c\,d$;
however, this assumption is not essential to the analysis.
For a more complete discussion of the real number model of computation
with function values, see \cite{N95,TWW88}. 

Let $A(f)$ be any algorithm for approximation of path integrals. 
The algorithm $A$ uses a finite number $n_1$ of function values at 
points $x_i$ and a finite number $n_2$ of arithmetic operations
and comparisons to compute $A(f)$.  The cost of computing $A(f)$ is
$\c_dn_1+n_2$. 
In the {\it worst case} setting, the error and cost of $A$ are defined 
by its worst performance over the class $F$. 
In the {\it randomized} setting, the algorithm $A$ may use
randomly chosen samples $x_i$, and its error and cost
are defined by the expected error with respect
to the distribution generating the random samples for a worst $f$ from $F$.
By the worst case or randomized {\it complexity}, we mean the minimal cost 
that is needed to compute an $\e$-approximation for all $f\in F$,
see \cite{TWW88} for precise definitions. 

We now briefly discuss the worst case and randomized complexities 
of path integration for the classes $F_r$.
We begin with the worst case setting. 
We first state the result of Bakhvalov, see e.g., \cite{N88,TWW88},
which states that the worst case complexity of multivariate 
integration over the unit cube $[0,1]^d$ for $r$-times differentiable 
functions is of order $\c_d\e^{-d/r}$. For path integration $d$ is an
increasing function of $\e^{-1}$, 
and as shown in Theorem 1, it goes to infinity 
polynomially in $\e^{-1}$ as $\e$ goes zero. 
This suggest that the worst case complexity,
$\comp^{{\rm wor}}(\e,F_r)$,
of path integration in the class
$F_r$ is exponential\footnote{We follow a convention
of complexity theory that if the complexity growth is faster than polynomial
then we say it is exponential.} in $\e^{-1}$. A formal 
proof may be found in \cite{WW96} for any $r$,
and more precise complexity bounds in \cite{C00} for $r=1$.
Thus  path integration is {\it intractable} for the class $F_r$
in the worst case setting. This means that the cost of any 
algorithm for solving this problem must be exponential. Yet,
as we shall see,  such algorithms will be useful for 
quantum computation. We now sketch such an algorithm.

We first consider the case $r=1$ and then show that an easy modification
of the same  algorithm can be also used for $r\ge2$. 
We assume\footnote{The theta notation means
that there exist positive numbers $c_1$ and $c_2$ such that
$c_1j^{-k}\le \l_j\le c_2j^{-k}$ for all $j=1,2,\dots$.} that  
$\l_j=\Theta(j^{-k})$. 
{}From Theorem 1 we know that it is enough 
to compute 
an ${\e\over 2}$-approximation to the integral
$I_d(f)$ with $d=d(\e,F_1)$ given in Theorem 1. We have
\begin{equation}\label{1570}
I_d(f)\,=\,\int_{\reals^d}f_d(t)\,\nu_d(dt),
\end{equation}
where $\nu_d$ is a Gaussian measure on $\reals^d$ with mean zero and
with the diagonal covariance matrix $\mbox{diag}(\l_1,\l_2,\dots,\l_d)$. 

This problem has been studied in \cite{C00}. Based on that paper
we describe an algorithm $S_n$ with worst case error at most
${\e\over 2}$. We opt here for simplicity of the presentation of $S_n$
at a slight expense of its cost. 
Let $n=m^d$ for the minimal odd  integer $m$ for which
\begin{equation}
m\,\ge\, \frac{4\,K_1\,\left(\pi\sum_{i=1}^d\l_i\right)^{1/2}}{\e}.
\end{equation}
For $x\ge0$, let 
$\psi(x)=\sqrt{2/\pi}\int_0^x\exp(-t^2/2)dt$ be the probability
integral, and let $\psi^{-1}$ be its inverse. We note that
it is easy to compute $\psi^{-1}(t)$ numerically for any $t\in \reals$.
As in Lemma~1 of \cite{C00}, for $i=1,2,\dots,d$ define the points $t_{i,j}:$
\begin{eqnarray*}
t_{i,0}\,&=&\,-\infty,\\
t_{i,j}\,&=&\,(3\l_i)^{1/2}\,\psi^{-1}\left((j-1/2)/m\right),
\qquad j=1,2,\dots,m,\\
t_{i,m+1}\,&=&\,\infty.
\end{eqnarray*}
Then take $t^*_{i,j}=t_{i,j}$ if $|t_{i,j}|\le |t_{i,j+1}|$,
and $t^*_{i,j}=t_{i,j+1}$ otherwise. For the integer vector
$\vec j=[j_1,j_2,\dots,j_d]$, with $j_i=1,2,\dots,m$, define the
sample points 
$$
x_{\vec j}\,=\,[t^*_{1,j_1},t^*_{2,j_2},\dots,t^*_{d,j_d}].
$$
Then the algorithm takes the simple form
\begin{equation}\label{alg}
S_n(f_d)\,=\,n^{-1}\,\sum_{{\vec j}} f\left(x_{\vec j }\right).
\end{equation}
Curbera proved in \cite{C00} that the worst case error of $S_n$
is at most 
$$
e^{{\rm wor}}\left(S_n\right)\,\le\, 2\,K_1\,\left(\pi\,\sum_{i=1}^d\l_i
\right)^{1/2}\,m^{-1} \,\le \,{\e\over 2},
$$
where the last inequality holds due to the choice of $m$. 

The cost of $S_n$ is $(\c_d+1)n$ where
\begin{equation}\label{n1}
n\,=\,n(\e^{-1})\,=\,
m^d\,\le \left(\frac{\a}{\e}\right)^{\overline{c_1}\e^{-2/(k-1)}},
\end{equation}
where $\a=4\,K_1\,\left(\pi\,\sum_{i=1}^\infty\l_i\right)^{1/2}+2\e$, 
and $\overline{c_1}$ is from Theorem 1.

We now consider the case $r\ge 2$. 
As shown in Theorem 1, we can now restrict ourselves to the integrals
$I_d(f)$ for $d=d(\e,F_r)$. We stress that $d(\e,F_r)$ 
is much less than $d(\e,F_1)$ for small $\e$. 
Observe that all functions $f$ from 
$F_r$ also belong to $F_1$ since they satisfy the Lipschitz condition with 
the constant $K_1$. Hence we can use the algorithm $S_n$ with the important
difference that now $d=d(\e,F_r)$. Hence, we compute an 
${\e\over 2}$-approximation
by the algorithm $S_n$ with cost $(\c_d+1)n$, where
\begin{equation}\label{n2}
n\,=\,n(\e^{-1})\,=\,m^{d(\e,F_r)}\,\le\,
\left(\frac{\a}{\e}\right)^{\overline{c_2}\e^{-1/(k-1)}},
\end{equation}
with $\overline{c_2}$ from Theorem 1. 

We now justify why it is enough to apply the algorithm $S_n$ 
for the class $F_r$ for any $r\ge2$. The reason is that 
for path integration the smoothness parameter $r$ 
is not as important as for finite dimensional integration. 
Indeed, since the exponent of $\e^{-1}$ for the worst case
complexity of path integration is unbounded (as $\e\to 0$) 
for any fixed $r$, it does not help much to divide by $r$. 

As we shall see in the next section, for quantum computation
the logarithm of the worst case complexity is important and 
$r$ can only effect a multiplicative factor. The most 
important property is how fast $d(\e,F_r)$ goes to infinity. As we know 
from Theorem 1, the influence of $r$ is significant here  since
we have different formulas for $d(\e,F_r)$ for $r=1$ and $r\ge2$. 
However, for $r\ge2$,
the use of more efficient algorithms than $S_n$ can only improve
the multiplicative factor of the logarithm of the worst case complexity.

We turn to the randomized setting for $\l_i=\Theta(j^{-k})$.  
The randomized complexity, $\comp^{{\rm ran}}(\e,F_r)$, can be easily
obtained by applying results of Bakhvalov for finite dimensional
integration. The analysis in \cite{WW96} yields
\begin{equation}\label{rand0}
\comp^{{\rm ran}}(\e,F_r)\,=\, \Theta\left((\c_d+1)\,
\e^{-2(1+o(1))}\right) \qquad \mbox{as}\ \ \e\to 0,
\end{equation}
where $d=d(\e,F_r)$ and 
the factors in the $\Theta$ notation depend at most quadratically
on $K_0$ and $K_1$. Hence, in the randomized setting we have roughly
quadratic dependence in $\e^{-1}$ on the number of function values. 
If $\c_d=\c\,d$ then 
\begin{equation}\label{rand}
\comp^{{\rm ran}}(\e,F_r)\,=\, 
\Theta\left(\c\,\left(\frac1{\e}\right)^{\frac{1+\gamma(r)}{k-1}\,+\,
2\,+o(1)}\right)
\end{equation}
where $\gamma(1)=1$ and $\gamma(r)=0$ for $r\ge 2$. 

Hence, the randomized complexity of path integration depends
polynomially on $\e^{-1}$, and therefore the path integration 
problem is {\it tractable} in the randomized setting. In fact,
the upper bound can be achieved by the Monte Carlo algorithm
with randomized error at most~${\e\over 2}$ and with
the cost proportional to $\c_d\, K_0^2\e^{-2}$ randomized evaluations of
a function of $d=d(\e,F_r)$ variables, where  $d(\e,F_r)$ 
is given by Theorem 1. 

Note, however, that if $k$ goes to one then the degree of $\e^{-1}$
in the randomized complexity goes to infinity. The reason is that in this
case we have to compute function values of very 
many variables. On the other hand,
for the Wiener measure we have $k=2$, and the degree of $\e^{-1}$
is roughly $3+\gamma(r)$.

\section{Path Integration on a Quantum Computer}

We now analyze path integration on a quantum computer.
The idea behind solving path integration on a quantum computer is quite
simple. (However, the analysis is not so simple.) We will apply 
analogous techniques for other problems in future papers. 

Start with an algorithm  that  computes an $\e$-approximation 
to path integration in the {\it worst} case setting and that requires 
summation of the form of (\ref{sum}). 
We run this algorithm on a quantum computer using the 
quantum summation algorithm of Section 2. 
Obviously, $n$ is now a function of $\e^{-1}$. For 
path integration for the class $F_r$ we know that $n$ 
is an exponential function of $\e^{-1}$ and is bounded by (\ref{n1}) 
for $r=1$, and by (\ref{n2}) for $r\ge 2$. 
However, the exponential dependence on $\e^{-1}$
is now {\it not} so essential since the cost of the quantum 
summation algorithm $QS_n$
depends only logarithmically on $n$. Since 
$\log\,n$ is a polynomial
in $\e^{-1}$ we conclude that path integration on a quantum computer 
can be solved at cost {\it polynomial} in $\e^{-1}$. 
That is, intractability of path integration in the worst case setting
is {\it broken} on a quantum computer 
by the use of the quantum summation algorithm. 

For other intractable problems in the worst case setting for which
the worst case complexity can be achieved by summation of $n$ numbers, 
intractability will be broken  as long as $n$ is a single exponential
function of $\e^{-1}$, i.e., $n(\e^{-1})\le 2^{p(\e^{-1})}$
with $p$ being a polynomial. Then the quantum cost will be polynomial
in $\e^{-1}$, and the problem will be tractable
on a quantum computer. 
This idea will {\it not} work if $n(\e^{-1})$ is a double 
exponential function (or worse) 
of~$\e^{-1}$ since then the logarithm of $n(\e^{-1})$  
will be 
still an exponential function of $\e^{-1}$. 

We now provide details of this idea  for path integration 
for the class $F_r$ with eigenvalues $\l_j=\Theta(j^{-k})$.
We take the algorithm $S_n$ defined by (\ref{alg})
with $n$ given by (\ref{n1}) for $r=1$ and by (\ref{n2}) for $r\ge2$.
The algorithm $S_n$ 
already has the form (\ref{sum}) required by the summation algorithm.
However, the summands $f(x_{\vec j})$ are not necessarily in the
interval $[-1,1]$. The function $f$ belongs to $F_r$ and therefore
its values are bounded by $K_0$. Hence, it is enough to scale the
problem by running the quantum summation algorithm for
$y_{\vec j}=f(x_{\vec j})/K_0$, replace $\e$ by $\e/K_0$,
and multiply the computed result by $K_0$.  
The cost of an algorithm on a quantum computer using $m$ qubits 
is defined as on a classical computer with the cost of a quantum query
taken as $\c_d+m$ since $f(x_{\vec j})$'s are computed 
and $m$ qubits are processed by a quantum query. 
Using the results of quantum summation
from Section 2 applied for $S_n$ we obtain the following theorem.
\vskip 1pc
\begin{thm}\label{thm2}
Consider path integration for the class $F_r$ with the eigenvalues
$\l_j=\Theta(j^{-k})$. Using the quantum summation 
algorithm $QS_n$ to compute an $\e/K_0$-approximation to $S_n$, we compute an
$\e$-approximation for path integrals with probability at least
${3\over 4}$ and of order
\begin{eqnarray*}
\e^{-1}&\qquad& \mbox{quantum queries},\\
\e^{-((k+\gamma(r))/(k-1)}\,\log\,\e^{-1}
&\qquad& \mbox{quantum operations, }\\
\e^{-((1+\gamma(r))/(k-1)}\,\log\,\e^{-1}\, 
&\qquad& \mbox{qubits}, 
\end{eqnarray*}
where $\gamma(1)=1$ and $\gamma(r)=0$ for $r\ge 2$.
If $\c_d=\c\,d$, then the cost of this algorithm is of order
$$
\left(\frac1{\e}\right)^{\frac{1+\gamma(r)}{k-1}\,+1}
\left(\c\,+\,\log\,\frac1{\e}
\right).
$$

For the Wiener measure the results are more precise. The algorithm requires
(neglecting ceilings for simplicity) at most
\begin{eqnarray*}
2K_0\,\e^{-1}&\qquad& \mbox{quantum queries},\\
\frac{2K_0}{\e}\,d^{{\rm up}}(\e,F_r)\log \,\frac{16\,K_0\,K_1}
{\e}
&\qquad& \mbox{quantum operations, }\\
d^{{\rm up}}(\e,F_r)\,\log \,\frac{16\,K_0\,K_1}{\e}
&\qquad& \mbox{qubits}.
\end{eqnarray*}
If $\c_d=\c\,d$, then the cost of this algorithm is at most
$$
\frac{2K_0\,d^{{\rm up}}(\e,F_r)}{\e}\,\left(\c\,+\,
2\,\log\,\frac{16\,K_0\,K_1}{\e}\right).
$$
Here $d^{{\rm up}}(\e,F_r)$ is an upper bound on $d(\e,F_r)$ given by
$$
d^{{\rm up}}(\e,F_r)\,=\,
\bigg\lceil\left(\frac{K_0\beta(r)}{\pi^2\e}\right)^{1+\gamma(r)}\,+\,\frac12
\bigg \rceil\,
$$
with  $\beta(r)=2K_1$ for $r=1$, and $\beta(r)=K_2$ for $r\ge2$. 
\end{thm}
\vskip 1pc
If we want to increase the probability of computing an $\e$-approximation
to path integration then, as explained in Section 2, 
we can run the quantum algorithm for $QS_n$ roughly
$\log\,\delta^{-1}$ times and take the median as the final result.
Then the probability of success is at least $1-\delta$.
Obviously the cost is then multiplied by $\log\,\delta^{-1}$ but the number
of qubits stays the same. 

We compare ${\rm cost}(QS_n)$, the cost of the quantum algorithm, with
the worst case complexity of path integration. The essence of Theorem 2 is 
that ${\rm cost}(QS_n)$ depends polynomially on $\e^{-1}$.
Since the worst case complexity is exponential in $\e^{-1}$, 
the use of quantum summation breaks intractability of the worst case setting. 
Note that we have {\it exponential speed-up}, 
i.e., ${\rm comp}^{{\rm wor}}(\e,F_r)/
{\rm cost}(QS_n)$ is exponential in $\e^{-1}$.

We now compare ${\rm cost}(QS_n)$ 
with the randomized complexity
of path integration. As discussed in Section 5, path integration is tractable
in the randomized setting and its randomized complexity is characterized
by (\ref{rand0}) and (\ref{rand}). Comparing the formulas for
the randomized complexity with ${\rm cost}(QS_n)$ we see that
the ratio of the number of quantum queries used by the
quantum algorithms to the number of function values used by
the best randomized algorithm is roughly $\e^{-1}$. 
If we compare ${\rm cost}(QS_n)$ to the randomized complexity
we see that the speed-up is roughly of order $\e^{-1}$. That is, we solve
path integration on a quantum computer roughly $\e^{-1}$ times cheaper than
on a classical computer using randomization. We summarize our results
in the following corollary.
\vskip 1pc
\begin{cor}
Consider path integration for the class $F_r$ with $\l_j=\Theta(j^{-k})$.
Then
\begin{itemize}

\item Path integration on a quantum computer is tractable.

\item Path integration on a quantum computer can be solved roughly
      $\e^{-1}$ times faster than on a classical computer using 
      randomization, and exponentially faster than on a classical computer
      with a worst case assurance. 

\item The number of quantum queries is the square root of 
      the number of function values needed on a classical
      computer using randomization. 

\item The number of qubits is polynomial in $\e^{-1}$.
      Furthermore, for the Wiener measure the degree is $2$ for $r=1$,
      and $1$ for $r\ge2$. 
\end{itemize}
\end{cor}

\section{Lower Bounds on the Number of Quantum Queries}

We now study lower bounds on the minimal number $\qq$ of quantum
queries needed to compute an $\e$-approximation with
probability $3/4$ for path integration for the class~$F_r$. 
{}From Theorem 2 we know that $\qq$ is at most of order $\e^{-1}$.
We show that this bound cannot be significantly improved. 

\begin{thm}
Consider path integration for the class $F_r$ with all positive
eigenvalues $\l_i$. Then
$$
\lim_{\e\to 0}\,\e^{1-\a}\,\qq\ =\ \infty\qquad \forall\,\a\in (0,1).
$$
\end{thm}

\proof The proof consists of two steps. The first one is to reduce
path integration to a finite dimensional Gaussian integration which is 
no harder than the original problem. The second step is essentially
the same as in Novak's papers, see \cite{N88,N01}, and reduces the
finite dimensional Gaussian integration problem to summation for which
the lower bound of Nayak and Wu, see \cite{NW98}, applies. 

In the first step of the proof, for a given $\a\in (0,1)$ we take an 
integer $d>r(1-\a)/\a$. (Hence, $d$ is large for small $\a$.)
The path integration problem for the class $F_r$ is no harder if we
assume some additional properties of functions $f$ from $F_r$. We
have, see (\ref{int}), 
$$
I(f)\,=\,\int_Hf(\mbox{Im}^{-1}x)\,\nu(dx)
$$
where $x=\sum_{i=1}^\infty t_i\eta_i$. 

Let us now assume that $f$ depends only on the first $d$ components
$t_1,t_2,\dots,t_d$, and call this class $F_{r,d}$. Obviously,
$F_{r,d}$ is a subclass of $F_r$ and therefore path integration
for $F_{r,d}$ is no harder than for the class $F_r$. For the class
$F_{r,d}$, the path integration problem reduces to a finite dimensional
Gaussian integration problem. That is, for $f\in F_{r,d}$ we have
$$
I_d(f)\,=\,\int_{\reals^d}f_d(t)\,\nu_d(dt),
$$
where $\nu_d$ is the Gaussian measure given by (\ref{1570}), and
$f_d$ is given by (\ref{1580}).  

The functions $f_d$ from $F_{r,d}$ are $r-1$ times 
continuously differentiable and 
$$\|f_d\|=\sup_{t\in\reals^d}|f_d(t)|\,\le\, K_0.
$$
Furthermore, their $r-1$ partial
derivatives satisfy the Lipschitz condition. More precisely, there
exists a positive number $\beta_1$ depending only on $d,r$ and
$K_0,K_1,\dots,K_r$ such that
$$
\left|D^if_d(t)-D^if_d(y)\right|\,\le\,\beta_1\|t-y\|_{\infty}\quad
\forall\,t,y\in \reals^d,
$$
where $D^i$ runs through the set of all partial derivatives of order
$r-1$. 

This shows that the class $F_{r,d}$ is closely related to the class
$F^{r-1,1}_d$ studied by Novak, see \cite{N01},
$$
F^{r-1,1}_d\,=\,\{f:[0,1]^d\to\reals\,|\, f\in C^{r-1}([0,1]^d),\,
|D^if(t)-D^if(y)|\le\|t-y\|_{\infty}\,\forall t,y\,\}.
$$
Obviously, the different Lipschitz constants: $\beta_1$ in our case
and $1$ for the class $F^{r-1,1}_d$ do not play a major role since
they do not change the order of error bounds.  
One difference between the two classes is that the common 
domain of functions from 
$F_{r,d}$ is $\reals^d$, whereas for the class $F^{r-1,1}_d$ the common
domain is $[0,1]^d$. A second difference is that we have Gaussian
integration whereas Novak considered uniform integration,
$\int_{[0,1]^d}f(t)\,dt$. As we shall see below these two
differences are not really essential.  

In the second step of the proof, we use Novak's proof technique. {}From
\cite{N88,N01} we know that for any positive $\e_1$ there are
functions $f_1,f_2,\dots f_n$, with $n=\Theta(\e_1^{-d/r})$, from the
class $F^{r-1,1}_d$ such that they take non-negative values, have
disjoint supports in $[0,1]^d$ and 
\begin{itemize}
\item $\int_{[0,1]^d}f_i(t)\,dt\,=\,\e_1^{1+d/r}\quad
  i=1,2,\dots,n$,
\item $\sum_{i=1}^n\a_if_i\,\in\,F^{r-1,1}_d\quad
  \forall\,|\a_i|\le 1$. 
\end{itemize}

We use the same functions $f_i$ for our Gaussian integration for the
class $F_{r,d}$. Since the support of $f_i$ is in $[0,1]^d$ we can
extend $f_i$ by zero to $\reals^d$. The extended functions $f_i$ have
exactly the same smoothness as required for the class $F_{r,d}$, and
there exists a positive $\beta_2$ depending only on $d,r$ and
$K_0,K_1,\dots, K_r$ but independent of $i$ such that $\beta_2f_i\in
F_{r,d}$. Note that
$$
\int_{\reals^d}f_i(t)\,\nu_d(dt)\,=\,
\int_{[0,1]^d}\rho_d(t)\,f_i(t)\,dt,
$$
where
$$
\rho_d(t)\,=\, \frac{1}{\prod_{j=1}^d(2\pi\l_j)^{1/2}}
\exp\left(-\sum_{j=1}^dt_j^2/2\right).
$$
Since all $\l_j$ are positive, the function $\rho_d$ has positive
minimum and maximum over $[0,1]^d$. That is, there are positive 
$\beta_3$ and $\beta_4$ depending on $d$ and $\l_1,\l_2,\dots,\l_d$
such that
$$
\beta_3\,\le\,\rho_d(t)\,\le \beta_4\quad \forall\,t_j\in[0,1].
$$
Therefore for the functions $g_i=\beta_2f_i\in F_{r,d}$ we have
\begin{equation}\label{ostatnia}
\beta_2\beta_3\,\e^{1+d/r}\,\le\,
I(g_i)\,=\,\int_{\reals^d}g_i(t)\,\nu_d(dt)\,\le\,\beta_2\beta_4
\e_1^{1+d/r},
\end{equation}
and 
$$
\sum_{i=1}^n\a_ig_i\,\in\,F_{r,d}\quad \forall\,|\a_i|\le 1.
$$
Since
$$
I\left(\sum_{i=1}^n\a_ig_i\right)\,=\,\sum_{i=1}^n\a_iI(g_i)
$$
we reduce our problem to summation of $n=\Theta(\e_1^{-d/r})$ terms
for arbitrary $|\a_i|\le 1$. Let
$$
y_i\,=\,\frac{I(g_i)}{\beta_2\beta_4\e_1^{1+d/r}}\,\a_i.
$$
Observe that by varying $\a_i$ from $[-1,1]$, the $y_i$ can take any
value from $\beta_3/\beta_4[-1,1]$ due to the
left hand side of (\ref{ostatnia}). 

We need to compute an $\e$-approximation to $I(\sum_{i=1}^n\a_ig_i)$.
This is equivalent to computing an $\e_2$-approximation to
$$
\frac1n\,\sum_{i=1}^ny_i
$$
with $\e_2=\e/(n\beta_2\beta_4\e_1^{1+d/r})=\Theta(\e/\e_1)$. 

The summation problem $n^{-1}\sum_{i=1}^ny_i$ for our $y_i$ is not
easier than the summation problem $n^{-1}\sum_{i=1}^ny_i$ for
all $|y_i|\le\beta_3/\beta_4$. We can now apply the lower bound of 
Nayak and Wu, see \cite{NW98}, that states that the minimal number of 
quantum queries needed to compute an $\e_2$-approximation with
probability $3/4$ for the summation problem $n^{-1}\sum_{i=1}^ny_i$
with $|y_i|\le \beta_3/\beta_4$ is bounded from below by
$$
C\,\min\left(n,\frac{\beta_4}{\beta_3\,\e_2}\right)\,=\,\Theta\left(
\min(\e_1^{-d/r},\e_1/\e)\right),
$$
with some absolute positive number $C$. 

Finally, we take $\e_1$ such that $\e_1^{-d/r}=\e_1/\e$, i.e., 
$\e_1=\e^{r/(d+r)}$, and conclude that the minimal number $\qq$
of quantum queries is at least of order $\e^{-(1-r/(d+r))}$. 
Since $d>r(1-\a)/\a$ implies that $\a>r/(d+r)$, we have
$$
\e^{1-\a}\,\qq\,=\,\Omega\left(\e^{-(\a-r/(d+r))}\right)\,
\to\,\infty \quad\mbox{as}\ \e\to 0.
$$
This completes the proof. \qed

\section{Appendix}

We prove Theorem 1.  We begin with $r=1$.
It is shown in \cite{WW96} that $d=d(\e,F_{1})\le d^*$ where $d^*$ is 
an integer satisfying 
\[
\sum_{i=d^*+1}^\infty\l_i\,\le\,\e^2/(2K_1)^2.
\]
For $\l_i\,=\,\Theta(i^{-k})$ with $k>1$, we get
\[
d^*\,=\,\Theta\left((K_1/\e)^{2/(k-1)}\right) \quad\mbox{as}\ \e\to0.
\]
For $\l_i=ai^{-k}$, we get 
$$
d^*\,=\,\bigg\lceil \left(\frac{a}{k-1}\right)^{1/(k-1)}
\left(\frac{2K_1}{\e}\right)^{2/(k-1)}\bigg\rceil.
$$
For the Wiener measure we have 
$$
d^*\,=\, \bigg\lceil\left(\frac1{\pi^2}\, 
\frac{2K_1}{\e}\right)^2\,+\frac12\bigg\rceil.
$$
This establishes upper bounds on $d(\e,F_1)$.

To get a lower bound, take the function $g:\reals\to \reals$ defined
by $g(x)=c_1|x|/(1+|x|)$ with $c_1=\min(K_0,K_1)$. We have 
$\sup_{x\in \reals}|g(x)|=c_1\le K_0$, and $g$ satisfies the Lipschitz
condition with the constant $c_1\le K_1$. For $x\in H$, define
$x_d=\sum_{j=d+1}^{2d}\il x,\eta_j\ir_H\eta_j$ and
$$
f\left({\rm Im}^{-1}x\right)\,=\,g(\|x_d\|). 
$$
Then $f\in F_1$ and $f_d=0$. We have $I(f)-I(f_d)=I(f)$ and 
$$
I(f)\,=\,\int_Hg(\|x_d\|)\,\nu(dx)\,\ge\,
c_1\int_{\|x_d\|\le1}\frac{\|x_d\|}{1+\|x_d\|}\,\nu(dx)\,\ge\,
\frac{c_1}2\int_{\|x_d\|\le1}\|x_d\|\nu(dx).
$$
Since $\left(\sum_{j=d+1}^{2d}a_j^2\right)^{1/2}\ge d^{-1/2}
\sum_{j=d+1}^{2d}|a_j|$ for any $a_j\in \reals$, we get
$$
I(f)\,\ge\,\frac{c_1}{ 2d^{1/2}}\, \sum_{j=d+1}^{2d}{\rm Int}_j
$$
where
$$
{\rm Int}_j\,=\,\prod_{i=1}^d(2\pi\l_i)^{-1/2}\int_{\sum_{i=d+1}^{2d}x_i^2\le1}
|x_j|\exp\left(-\sum_{i=d+1}^{2d}x_i^2/(2\l_i)\right)\,dx.
$$

There exist two positive numbers $\a_1$ and $\a_2$ such that
$\a_1i^{-k}\le\l_i\le\a_2i^{-k}$ for all $i$. 
By changing variables $t_{i-d}=x_i/(\l_i)^{1/2}$ and noting that 
$\a_2d^{-k}\ge\l_d\ge \l_i\ge\l_{2d}\ge\a_1(2d)^{-k}$ we conclude that
\begin{eqnarray*}
{\rm Int}_j\,&=&\,
\frac{\sqrt{\l_j}}{(2\pi)^{d/2}}
\int_{\sum_{i=1}^{d}\l_it_i^2\le1}
|t_{j-d}|\exp\left(-\sum_{i=1}^{d}t_i^2/2\right)\,dx,\\
{\rm Int}_j\,&\ge&\,\frac{\a_1^{1/2}}{(2d)^{k/2}}\left(
\frac1{(2\pi)^{d/2}}\int_{\reals^d}|t_1|\,e^{-\|t\|^2/2}\,dt\,-\,
\frac1{(2\pi)^{d/2}}\int_{\|t\|>(d^k/\a_2)^{1/2}}
|t_1|\,e^{-\|t\|^2/2}\,dt\right).
\end{eqnarray*}
Here, $t=[t_1,t_2,\dots,t_d]$ and $\|t\|=(\sum_{j=1}^dt_j^2)^{1/2}$.
The first integral is just $\sqrt{2/\pi}$. We now show that the second
integral goes to zero with $d$. Indeed, let
$\nu_d$ be for the Gaussian measure on $\reals^d$ with zero mean
and the identity covariance operator, and let $\alpha=(d^k/\a_2)^{1/2}$.
Then the second integral is $\int_{\reals^d\setminus B_\alpha}|t_1|\nu_d(dt)$,
where $B_\alpha$ denotes the ball of radius $\alpha$, and is 
not greater than
$$
\left(\int_{\reals^d}t_1^2\nu_d(dt)\right)^{1/2}\left(1-\nu_d(B_\alpha)
\right)^{1/2}.
$$
The integral with the integrand $t_1^2$ is just one, and using Lemma 2.9.2
from \cite{TWW88} p.~469 we conclude that
$$
1-\nu_d(B_\alpha)\,\le\, 5 \exp\left(-\alpha^2/(2d)\right).
$$
Since $\alpha^2/(2d)=d^{k-1}\a_2/2$ and $k>1$ then this ratio goes to 
infinity, and $1-\nu_d(B_d)$ goes to zero.
This means that ${\rm Int}_j$ is at least of order $d^{-k/2}$, and $I(f)$ is
at least of order $d^{-(k-1)/2}$. Hence to guarantee that $I(f)=I(f)-I(f_d)\le
{\e\over 2}$ we must take $d$ of order $\e^{-2/(k-1)}$ which completes the proof
for the case $r=1$.

Assume now that $r\ge 2$. We first establish an upper bound on $d(\e,F_r)$. 
For $x=\sum_{j=1}^\infty t_j\eta_j\in H$, define
$t=[t_1,t_2,\dots]\in \reals^\infty$ and $t^d=[t_1,t_2,\dots,t_d]$.
Then we can identify $f(t)$ with $f({\rm Im}^{-1}x)$ and $f(t^d)$ with
$f_d(t)$. By Taylor's theorem we have
$$
f(t)\,=\,f(t^d)+f^{\prime}(t^d)(t-t^d)+
\int_0^1\left(
f^{\prime}\left(t^d+u(t-t^d)\right)-f^{\prime}(t^d)\right)(t-t^d)\,du.
$$
Note that $t-t^d=[0,\dots,0,t_{d+1},t_{d+2},\dots]$ and since 
$f^{\prime}(t^d)$ is a linear form we have 
$$
f^{\prime}(t^d)(t-t^d)\,=\,\sum_{j=1}^\infty a_jt_j,
$$
where $a_j=a_j(t_1,t_2,\dots,t_d)$. The mean element of $\nu$ is zero, 
which implies that $I(f^{\prime}(t^d)(t-t^d))=0$. Hence
$$
I(f)-I(f_d)\,=\, 
I\left(\int_0^1\left(f^{\prime}\left(t^d+u(t-t^d)\right)-
f^{\prime}(t^d)\right)(t-t^d)\,du\right).
$$
For $r\ge2$, $f^{\prime}$ satisfies the Lipschitz condition and we get
\begin{eqnarray*}
\left|I(f)-I_d(f)\right|\,&\le&\,K_2\,I\left(\|t-t^d\|^2\,\int_0^1u\,du
\right)\\
&\le&\,\frac{K_2}2\,I\left(\sum_{j=d+1}^\infty\il x,\eta_j\ir^2_H\right)\,=\,
\frac{K_2}2 \sum_{j=d+1}^\infty\l_j.
\end{eqnarray*}
For $\l_j=\Theta(j^{-k})$ we obtain
$$
|I(f)-I_d(f)|\,=\,O\left(d^{-(k-1)}\right)
$$
and for $d=O(\e^{-1/(k-1)})$ 
we guarantee that $|I(f)-I(f_d)|\le {\e\over 2}$ for all $f\in F_r$.

For $\l_j=aj^{-k}$, we have 
$$
\sum_{j=d+1}^\infty\l_j\,\le\, a\,\int_d^\infty u^{-k}\,du\,=\,
\frac{a}{2(k-1)}\ \frac1{d^{k-1}}.
$$
In this case it is enough to take
$$
d\,=\,\bigg \lceil \left(\frac{aK_2}{k-1}\right)^{1/(k-1)}\,
\left(\frac1{\e}\right)^{1/(k-1)}\bigg \rceil.
$$
For the Wiener measure,
$$
\sum_{j=d+1}^\infty\l_j\,\le\, \frac1{\pi^2}\,\int_d^\infty 
\left(u-1/2\right)^{-2}\,du\,=\, \frac1{\pi^2(d-1/2)},
$$
and
$$
d\,=\, \bigg \lceil \frac{K_2}{\pi^2\e}\,+\,\frac12\bigg \rceil.
$$
This establishes upper bounds on $d(\e,F_r)$. 

To get a lower bound, consider the function $g(x)=c_rx^2/(1+x^2)=
c_r(1-1/(1+x^2))$ for $x\in \reals$, where $c_r$ is a positive number 
chosen such that
$\max_{0\le i\le r-1}\sup_{x\in \reals}|g^{(i)}(x)|/K_i\le1$,
and such that
$g^{(r-1)}$ satisfies the Lipschitz condition with the constant $K_r$. 
It is easy to see that such a positive number $c_r$ exists. Indeed, 
the $j$th derivatives of $1/(1+x^2)$ can be written 
as the ratio of two polynomials
$p_j(x)/(1+x^2)^{j+1}$ with the degree of $p_j$ being at most $j$, 
and therefore all derivatives go to zero as $|x|$ goes to infinity. 

As for the case $r=1$, we take $x_d=\sum_{j=d+1}^{2d}\il x,\eta_j\ir_H\eta_j$,
and $f({\rm Im}^{-1}x)=g(\|x_d\|)$. Then $f\in F_r, f_d=0$ and
$I(f)-I(f_d)=I(f)$. Similarly as for $r=1$ we have
$$
I(f)\,\ge\,\frac{c_r\l_{2d}}2\,\int_{\|t\|\le \alpha}\|t\|^2\nu_d(dt),
$$
where $\alpha=(d^k/\a_2)^{1/2}$. We now show that the last integral 
tends to $d$. Indeed, it can be written as
$$
\int_{\reals^d}\|t\|^2\nu_d(dt) \,-\,\int_{\|t\|>\alpha}\|t\|^2\nu_d(dt).
$$
The first integral is obviously $d$, and we show that the integral
over the outside of the ball tends to zero. For large $d$, 
the norm of $t$ is also large, and we can estimate 
$\|t\|^2\le\exp(c_d\|t\|^2/2)$ for $c_d=2\a_2\ln(d^k/\a_2)/d^k$.
Then 
$$
\int_{\|t\|>\alpha}\|t\|^2\nu_d(dt)\,\le\, (1-c_d)^{d/2}\,
\int_{\|t\|>\alpha}\nu_{d,c}(dt),
$$
where $\nu_{d,c}$ is a Gaussian measure on $\reals^d$
with mean zero and covariance operator
$(1-c_d)^{-1}I$. Again using Lemma 2.9.2 from \cite{TWW88} we obtain
$$
\int_{\|t\|>\alpha}\|t\|^2\nu_d(dt)\,\le\, 5(1-c_d)^{d/2}
\exp\left(-\alpha^2(1-c_d)/d\right).
$$
Since $k>1$, the quantity $(1-c_d)^d$ tends to $1$, and since
$\alpha^2/d=\Theta(d^{k-1})$ tends to infinity, the integral goes to zero 
as claimed.

Hence, $I(f)$ is at least of order $d^{-(k-1)}$ and $d$ must be at least 
of order $\e^{-1/(k-1)}$ to guarantee $I(f)\le {\e\over 2}$. 
This completes the proof
for $r\ge2$. \  \ \qed.

\vskip 2pc
\section*{Acknowledgment}

We are grateful for the excellent facilities of the Santa Fe Institute 
where some of our research was conducted. The lower bound theorem
of Section 7 was obtained during a stay at Los Alamos National
Laboratory. We are also grateful to S. Heinrich, E. Novak, A. Papageorgiou,
G. W. Wasilkowski and A. G. Werschulz for valuable comments on our paper.

\end{document}